\documentclass{emulateapj}
\usepackage{graphicx}

\shorttitle{A Microglich in PSR B1821-24}
\shortauthors{Cognard \& Backer}

\begin{document}

\pagestyle{empty}

\title{A Micro-glitch in the Millisecond Pulsar B1821-24 in M28}

\author{Isma\"el Cognard}
\affil{LPCE-CNRS, Orleans 3A, Av de la Recherche Scientifique, F-45071 ORLEANS, CEDEX2 -FRANCE-}
\email{icognard@cnrs-orleans.fr}

\author{Donald C. Backer}
\affil{Astronomy Department, 601 Campbell Hall, University of California, Berkeley, CA 94720-3411 -USA- }
\email{dbacker@astro.berkeley.edu}


\begin{abstract}
We report on the observation of a very small glitch observed
for the first time in a millisecond pulsar, PSR B1821-24 located in the globular cluster M28. 
Timing observations were mainly conducted with the Nan\c cay radiotelescope (France)
and confirmation comes from the 140ft radiotelescope at Green Bank and the new Green Bank Telescope data.
This event is characterized by a rotation frequency step of 3 nHz, or $10^{-11}$ in fractional frequency change
along with a short duration limited to a few days or a week.
A marginally significant frequency derivative step was also found.
This glitch follows the main characteristics of those in the slow period pulsars, 
but is two orders of magnitude smaller than the smallest ever recorded.
Such an event must be very rare for millisecond pulsars since no other glitches have been detected
when the cumulated number of years of millisecond pulsar timing observations up to 2001
is around 500 for all these objects. However, pulsar PSR B1821-24
is one of the youngest among the old recycled ones and there is likely a correlation
between age, or a related parameter, and timing noise.
While this event happens on a much smaller scale, the required adjustment of the
star to a new equilibrium figure as it spins down is a likely common cause for
all glitches.
\end{abstract}



\section{Introduction}

In long period pulsars rotational irregularities are observed that are thought to result
from erratic internal structure changes.
Two types of irregularities are observed: ``timing noise'', which is 
a slow unpredictable fluctuation of rotation;  and ``glitches'',
which are sudden changes in the rotation rate.
Glitches mainly occur in younger pulsars (Lyne, Shemar \& Graham-Smith 2000), and
are usually interpreted as a sudden transfer of angular momentum
from a faster-rotating component of superfluid to the solid crust of the star
(Baym et al. 1969, Anderson \& Itoh 1975).
These phenomena provide an opportunity to study the internal structure of neutron stars.
Days or months are necessary to recover after the glitch and such long relaxation
have been interpreted as strong evidence for superfluid in the neutron star (Sauls 1989).
Glitches have been observed with fractional frequency step amplitudes 
down to the level of $10^{-9}$ by Shemar \& Lyne (1996). They further 
estimated that 90\% of the glitches with size greater than $5\times10^{-9}$ are detected.
Despite carefull inspection of many years of high precision timing results 
of millisecond pulsars by several groups, no glitches have been observed in this
subcategory of pulsars. This fact is consistent with the statistical result that
the oldest pulsars, like the millisecond pulsars, are relatively free from glitches
(Lyne, Shemar \& Graham-Smith 2000).
The Nan\c cay radiotelescope is involved in a long term timing program on two millisecond pulsars for
more than 15 years. 
In this Letter, we report the first micro-glitch observed in a recycled millisecond pulsar which is
two orders of magnitude lower than any previously detected.


\section{Observations}

\begin{figure*}
\plotone{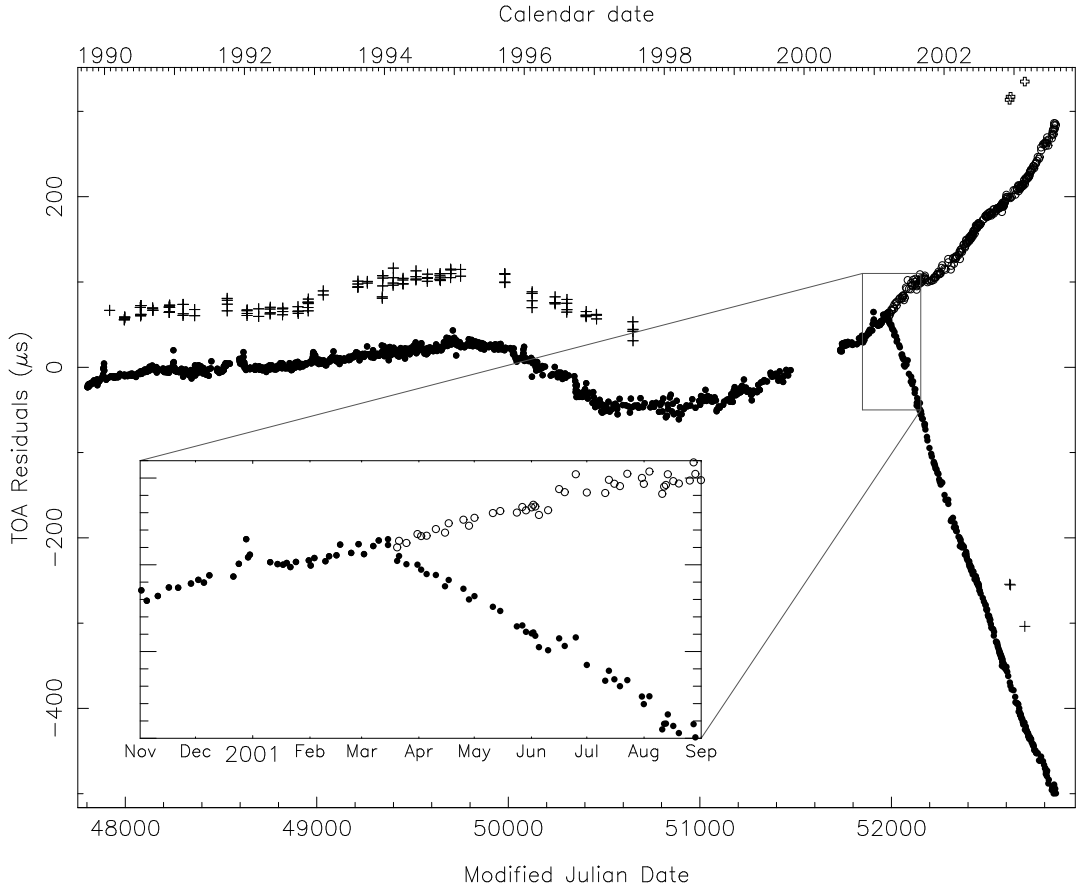}
\caption{
Timing residuals of PSR1821-24 at Nan\c cay Radio Observatory from October 1989 to August 2003
around radio frequency 1.4 GHz. Crux markers from 1989 to 1997 and in late 2002 and early 2003
are Green Bank TOAs.
A set of parameters (period, period derivative, position, proper motion plus few minors ones)
is adjusted over the Oct 1989 to Feb 2001 period and used to minimize the TOA residuals.
After March 2001, both TOAs from original set of parameters and TOAs obtained
with a $\delta$P/P change of $10^{-11}$ are shown.
The inset is a enlargement of the transition region showing that this transition does
not last more than a few days or a week. Also clearly visible is the effect of the extra
dispersive delay due to the occultation by the solar corona occuring each year around
Christmas (Cognard et al., 1996).
\label{fig1}}
\end{figure*}

The timing observations presented here have been conducted at the large decimetric radio
telescope (7000 square-meters, equivalent to a 93m dish) located near Nan\c cay (France).
We have gathered observations of PSR1821-24 since 1989 October about 10 times per month at 1.4 GHz.
At Nan\c cay, the pulsar signal is coherently de-dispersed prior to detection by using
a swept frequency local oscillator (first at 80 MHz and then at 150 MHz since 2000) in the IF chain.
The pulse spectra are produced by the station digital autocorrelator.
While most timing observations use a fixed center frequency and determine 
a pulse arrival time at this frequency, our swept-LO approach leads us to determine
a frequency of arrival for the start time of the observation by a standard cross-correlation
of the daily integrated pulse with a pulse template (Cognard et al. 1996).
The Nan\c cay observations, which are occasionally done at 1.68 GHz to monitor the variations 
of the column density of dispersing electrons,
were suspended between 1999 November and 2000 July owing to the upgrade of the telescope.
Observations of B1821-24 were also conducted with the 140ft Green Bank telescope between 1989 and 
1997, and briefly with the new 100m Green Bank Telescope (GBT) during 2002 and 2003.

The analysis of the pulsar timing data has been carried out by our software package (AnTiOPE)
that fits the observations to selected parameters involving the pulsar spin and astrometry
along with delays due to dispersive effects (Cognard et al. 1995).
The Earth's orbital motion is obtained from the Jet Propulsion Laboratory ephemerides DE407 and 
the adopted timescale is International Atomic Time (TAI). 
If a model for the Times Of Arrival (TOAs) is complete, then the residuals, observed TOAs minus 
model TOAs, will have a Gaussian distribution about zero. 
A few months after 2001 March, we found that the standard model of B1821-24 was inadequate.
Figure 1 shows the timing residuals based on a model up to 2001 March and the abrupt
change in slope -- a glitch in the rotation frequency.
Various checks were done regarding the instrumentation. The most convincing 
evidence for a real effect came from the lack of any similar effect in the
analysis of timing observations of other millisecond pulsars (MSPs) in the Nan\c cay program.
We also checked that the effect was not a function of the observing radio 
frequency and therefore could not be the result of 
dispersion measure variations (Cognard \& Lestrade 1997). 
Moreover, the Green Bank measurements, with two different telescopes before and after 2001 March
but at the same frequency, provide confirming evidence (Fig. 1).


\section{Results}

The parameters of a glitch in rotation 
(phase step, permanent frequency step, permanent frequency derivative step)
were added to AnTiOPE so that we could obtain a consistent overall fit. 
Table 1 shows the different fitted parameters. We tried to fit for a phase step at the 
date of the glitch, but nothing reliable was found. A step of the order of 15 $\mu$s was 
found, but at the price of a very clear discontinuity in the TOA residuals. We decided not 
to include this parameter in the fit.
A frequency derivative step was also found, but its significance was marginal and the presence
of timing noise is a significant source of systematic error.
The short duration of the event (inset of Fig.1), which is limited to less than a week, is 
the most important fact that leads to our glitch interpretation, as opposed to timing noise 
as discussed below.

Because of the dramatic change seen, we are driven to also explore the pulsar spin frequency variability
by binning data before and after the glitch.
A robust approach was then chosen with independent fits over
different sub-sets of data. Fits for frequency (and arbitrary rotational phase) were done
over data sets with durations of 100 and 365 days.
The position and proper motion parameters from the
whole data set were used to fix the position in each data subset.
The results in Figure 2 show a frequency step 
of approximately 3 nHz, or $10^{-11}$ in fractional frequency change, occuring in March 2001.
Note that in both cases, 100 days or 365 days, the data bins were phased with the date of 2001 March 10
when the event is assumed to have occured. 
Despite some obvious noise in the rotational frequency (e.g., during 1996) the frequency step is confirmed.
The frequency residuals could be further characterized as showing a glitch in frequency derivative
in late 1996 with magnitude of 0.5 nHz y$^{-1}$, or $1.7\times 10^{-17}$ s$^{-2}$ ;
we will not discuss this possibility further in this short account.
However, note the long term nature of this episode contrasting
with the sudden frequency change of the 2001 glitch.

\begin{figure}
\plotone{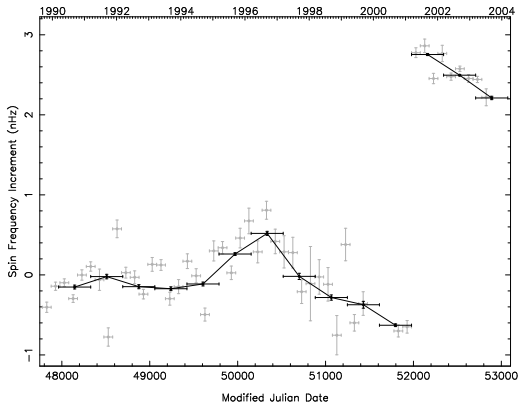}
\caption{
Evolution of the PSR B1821-24 rotational frequency. Variations around a mean value are plotted in nanoHertz
as a function of the date.
Black points and line are for frequencies determined over a full year (365days) while gray points were obtained
over shorter data spans of only 100days.
Horizontal error bars indicate the duration of the box used to derive the value.
The glitch occuring in March 2001 is clearly visible as a step of the order of 3nHz.
\label{fig2}}
\end{figure}

The 2001 March frequency glitch is two orders of magnitudes lower than any previously observed.
The smallest ones were observed at the level of $10^{-9}$ in pulsars: 0525+21 (Downs 1982), 
1907+00 (Gullahorn et al. 1976) and 0740-28 (D'Alessandro et al. 1993). However discrimination
between glitches and timing noise is difficult at this level. Shemar \& Lyne (1996)
estimated that 90\% of the glitches with relative frequency step greater than $5\times 10^{-9}$ were detected.
Here, we are observing a millisecond pulsar with a typical TOA uncertainty of 2 $\mu$s, which 
together with a high timing precision and stability allows easy detection of a glitch as small as $10^{-11}$.
This small event may indicate 
 a new regime in the evolution of neutron star internal structure. 
Indeed there are some very small glitches which may have been identified of both signs
in term of rotational frequency increment from the very young Vela pulsar (Cordes et al. 1988).
More events in B1821-24 and other MSPs are required to provide a clearer picture.
Part of the surprise and signifiance is that other MSPs are so stable.
PSR B1937+21 is known to be unstable at the microseconds level (Kaspi et al. 1994, Lommen \& Backer 2001),
but no glitches down to the $10^{-11}$ level have been seen in over 20 years. The same is true for many other
MSPs. We estimate that the cumulated number of years of millisecond pulsar timing observations up to 2001
is around 500 for all objects.

In order to assess the amount of potential timing noise present in B1821-24 data,
we performed a fit of harmonically related sinusoids to the timing residuals in the interval preceding
the 2001 March micro-glitch following the work of Hobbs (2002).
The rms of the curve produced can be seen as a measure of the timing noise present in the data.
The value obtained for B1821-24 is 24 $\mu$s. This value is a reasonable extrapolation of the
trend of rms vs period derivative $\dot P$ found by Hobbs (2002) and earlier authors. The value is larger than
but comparable to that of the 1.5ms pulsar B1937+21, but significantly larger than that
for other MSPs (Backer 2004).

In the known large glitches in slow pulsars, the sudden increase in rotational frequency is
followed by an exponential recovery and a possible permanent change in period derivative.
Glitches with exponential recovery observed in young pulsars indicate that the neutron
star is probably made of two components, crust and superfluid interior, that are
partially coupled together. 
A fraction of the quantized, superfluid vortices can be 
pinned to defects in the crust to trap angular momentum as the star spins down and
the vortices migrate outward
(Alpar et al. 1981). 
The trapped angular momentum is suddenly released at the time of a glitch
by an unexplained catastrophic unpinning event where angular momentum is transfered
from the inner part of the crust to the crust to produce the observed speed-up.

Glitch activity is thought to be maximum for young or adolescent pulsars, with ages up to 
$2\times 10^4$ y.
Ages of pulsars are usually estimated from the rotation evolution with a characteristic age
given by $P/2\dot P$ 
assuming a magnetic dipole braking index of 3 (Lyne \& Graham-Smith 1998).
Determination of the age of a pulsar in a globular cluster can be corrupted by
the gravitational potential through the period derivatives measurements.
However, Phinney (1993) showed that contribution from the cluster to the period
first derivative is limited to 6\% with a $\dot P$ term of $0.3^{-16}$ s$^{-1}$ at maximum. 
The characteristic age of $30\times10^6$ years for B1821-24 is therefore
fairly well established, and makes this millisecond pulsar the youngest among the 
globular cluster and galactic field MSPs.

\begin{table}
\caption[]{
PSR B1821-24 parameters obtained with analysis software AnTiOPE, TAI time scale
and JPL-DE407 ephemeris.}
\begin{tabular}{ll}
\hline
Spin and astrometry parameters & \\
\hline
Frequency (Hz)                          &   327.405641011622(5) \\
Frequency 1st derivative (Hz/s)         &   -173.5288(1)e-15    \\
Frequency second derivative (Hz/s2)     &   -4.15(7)e-26        \\
Epoch                                   &   2451468.5           \\
Right Ascension J2000                   &   18 24 32.00821(3)   \\
Declination J2000                       &   -24 52 10.784(8)    \\
RA proper motion (mas/yr)               &   0.27(8)             \\
DEC proper motion (mas/yr)              &   -15.2(16)           \\
\hline
Glitch parameters & \\
\hline
Glitch Epoch                            &   2451980.0           \\
Glitch Frequency step (Hz)              &   3.12(3)e-9          \\
Glitch Frequency derivative step (Hz/s) &   5.9(5)e-18          \\
\end{tabular}
\end{table}

Cordes et al. (1988) find a bimodal distribution of discontinuities in the
timing record of the Vela pulsar. The large events are the conventional
``glitches'' showing sudden spinup and subsequent exponential decay, and
amplitudes ($|\Delta \nu / \nu|$) of $\sim 10^{-6}$. They call the small
events ``microjumps''. These have amplitudes of $\sim 10^{-9}$ and are
{\it not} merely scaled down versions of the glitches. They suggest that
random timing noise in other pulsars is caused by unresolved microjumps.
With just a single event in the ensemble of millisecond pulsar timing 
records, it is difficult to speculate on how the glitch, jump and timing
noise properties evolve with age and magnetic field. For example,
Lyne et al. (2000) say ``there is a broad range of glitch amplitude,
and there is a strong indication that pulsars with large magnetic fields
suffer many small glitches while others show a smaller number of large
glitches.'' Following this to millisecond pulsars with small magnetic
fields, they should exhibit small number of large glitches. One microglitch
does not fit in with this statement beyond suggesting that a wide
distribution of glitch amplitudes is possible. Glitch amplitude distribution
and timing noise (e.g., Backer 2004, Hobbs 2002) in the millisecond pulsar population
will give us more insight into the nature of neutron star
interiors in the future.


\section{Summary}

For the first time, a glitch was observed in a recycled millisecond pulsar.
We must note that this pulsar appears to be one of the younger MSPs.
Long term observations of recycled pulsars are needed to assess the effect
of this behavior and influence on long term studies like
the search for background of primordial gravitational waves.
If other glitches are detected in very stable recycled pulsars, the perspective
of such long term goals will need to be thought again.

\acknowledgments
I.C. thanks G.Theureau and R.W.Hellings for fruitful discussions and
the support of the Nan\c cay Radio Observatory which is part of the Paris Observatory,
associated with the French Centre National de la Recherche Scientifique (CNRS).
The Nan\c cay Observatory also gratefully acknowledges the financial support of
the Region Centre in France. D.B. acknowledges both the staff at the National Radio Astronomy Observatory
and NSF grants.



\begin{thebibliography}{}

\bibitem[Alpar et al.(1981)]{alp81} Alpar, M.A., Anderson, P.W., Pines, D., Shaham, J,
1981, \apjl, 249, L29

\bibitem[Anderson & Itoh(1975)]{and75} Anderson, P.W., \& Itoh, N. 1975, {\it Nature}, {\bf 256}, 25

\bibitem[Backer 2004]{backer04} Backer, D.C., 2004, Aspen Conference on Binary Pulsars, ed. F. Rasio and I. Stairs
[ASP : San Francisco] (in press).

\bibitem[Baym et al. 1969]{bay69} Baym, G., Pethick, C., Pines, D., Ruerman, M., 1969, \nat, 224, 872
 
\bibitem[Cognard et al.1995]{cog95} Cognard, I., Bourgois, G., Lestrade, J.-F., Biraud, F., Aubry, D., Darchy, B.,
1995, \aap, 296, 169

\bibitem[Cognard et al.(1996)]{cog96} Cognard, I., Bourgois, G., Lestrade, J.-F., Biraud, F., Aubry, D., Darchy, B.,
1996, \aap,  311, 179

\bibitem[Cognard and Lestrade(1997)]{cog97} Cognard, I., Lestrade, J.-F.,
1997, \aap, 323, 211

\bibitem[Cordes, Downs and Krause-Polstorff(1988)]{dow88} Cordes, J.M., Downs, G.S., Krause-Polstorff, J.,
1988, \apj, 330, 847

\bibitem[D'Alessandro et al.(1993)]{dal93} D'Alessandro, F., McCulloch, P.M., King, E.A., Hamilton, P.A., McDonnell, D.,
1993, \mnras, 261, 883

\bibitem[Downes(1982)]{dow82} Downes, G.S., 1982, \apjl, 257, L67

\bibitem[Gullahorn et al.(1976)]{gul76} Gullahorn, G.E., Payne, R.R., Rankin, J.M., Richards, D.W.,
1976, \apjl, 205, L151

\bibitem[Hobbs (2002)]{hob02} Hobbs, G., 2002, PhD thesis Jodrell Bank Observatory, University of Manchester, UK

\bibitem[Kaspi, Taylor and Ryba(1994)]{kas94} Kaspi, V.M., Taylor, J.H., Ryba, M.,
1994, \apj, 428, 713

\bibitem[Lommen and Backer]{lom01} Lommen, A.N., Backer, D.C.,
2001, \apj, 562, 297

\bibitem[Lyne and Graham-Smith(1998)]{lyn98} Lyne, A.G., Graham-Smith, F., 1982, Pulsar Astronomy. Cambridge University Press

\bibitem[Lyne, Shemar and Graham-Smith(2000)]{lyn00} Lyne, A.G., Shemar, S.L., Graham-Smith, F.,
2000, \mnras, 315, 534

\bibitem[Phinney(1993)]{phi93} Phinney, E.S., 1993, ASP Conference Series, Vol. 50, 141

\bibitem[Sauls(1989)]{sau89} Sauls, J.A. 1989 in {\it Timing Neutron Stars}, edited by H. \"Ogelman
and E.P.J. van den Heuvel [Kluwer - Dordrecht] p. 457

\bibitem[Shemar and Lyne]{she96} Shemar, S.L., Lyne, A.G., 1996, \mnras, 282, 677

\end{thebibliography}
\end{document}